\documentclass[aps,pre,twocolumn,groupedaddress]{revtex4-1}
\usepackage{setspace}
\usepackage{graphicx}% Include figure files
\usepackage{subfigure}% Useful for putting images side-by-side
\usepackage{epsfig}
\usepackage{array}
\newcolumntype{M}{ >{\centering\arraybackslash} m{1cm}}
\newcolumntype{P}{ >{\centering\arraybackslash} m{3.85 cm}}
\usepackage{dcolumn}% Align table columns on decimal point
\usepackage{bm}% bold math
\usepackage{amsmath, amsthm, amssymb}
\usepackage[margin=1.0in]{geometry}
\usepackage{url}

\usepackage{xcolor}

\newcommand{\beq}{\begin{equation}}
\newcommand{\beqs}{\begin{equation*}}
\newcommand{\eeq}{\end{equation}}
\newcommand{\eeqs}{\end{equation*}}

\begin{document}

\title{Phase Separation and Emergent Structures in an Active Nematic}

\author{Elias Putzig}
\email[]{efputzig@brandeis.edu}
\author{Aparna Baskaran}
\email[]{aparna@brandeis.edu}
\affiliation{Martin Fisher School of Physics, Brandeis University, Waltham, MA 02453, USA}

\date{\today}

\begin{abstract}
We consider a phenomenological continuum theory for an active nematic fluid and show that there exists a universal, model independent instability which renders the homogeneous nematic state unstable to order fluctuations.  Using numerical and analytic tools we show that, in the vicinity of a critical point, this instability leads to a phase separated state in which the ordered regions form bands in which the direction of nematic order is perpendicular to the direction of density gradient.  We argue that the underlying mechanism that leads to this phase separation is a universal feature of active fluids of different symmetries.
\end{abstract}

\maketitle

\section{Introduction}

   Active fluids encompass diverse systems ranging from bacterial colonies \cite{Mendelson1998, Sliusarenko2006, Wu2000} to herds of animals \cite{Parrish1997} and bird flocks \cite{Ballerini2008}. These systems are unified by the fact that they are composed of ``microscopic" entities that consume energy and dissipate it to do work on their environment \cite{Ramaswamy2010,Marchetti2013,Vicsek2012}. Depending on the symmetries of the microscopic particles and the interactions among them, these systems can be classified as isotropic (ex: self-propelled spherical colloidal particles \cite{Palacci2013}), polar (ex: self-chemotactic bacteria \cite{Saragosti2011}) or nematic active fluids (ex: microtubule-motor-protein suspensions \cite{Sanchez2012, Ross2011}, vibrated granular rods \cite{Narayan2007}). In this work, we consider a minimal description of an active nematic fluid with the goal of identifying universal mechanisms for the formation of emergent structures on long length scales.

 Active nematics in general fall into two broad categories. The first is the self-propelled nematic, composed of self-propelled particles whose interactions have a nematic symmetry. This system has mixed symmetry in that the microscopic entity is polar (due to self-propulsion) but the interactions and therefore the macrodynamics is nematic and has been extensively studied in the literature \cite{Baskaran2008a,Baskaran2008, Ginelli2010, McCandlish2012,Yang2010,Peshkov2012,Baskaran2012,Farrell2012,Kraikivski2006,Peruani2006,Peruani2011a}. The second category is a pure active nematic composed of shakers, i.e., nematogens that do not undergo any persistent motion along either direction of their body axis. Physical realizations of pure active nematics include the microtubule suspensions mentioned above \cite{Sanchez2012}, symmetric vibrated rods \cite{Narayan2007}, rapidly reversing strains of myxobacteria \cite{Wu2011,Starruss2012} and melanocytes which are also thought to effectively behave as ``shakers" \cite{Duclos2014,Gruler2012,Kemkemer2000}. This latter class of active nematics are the focus of the study presented in this paper.

As with all realizations of active fluids, the microscopic entities that compose an active nematic fluid live in a medium (typically a viscous fluid) that acts as a momentum sink. When the flow induced by the active nematogens is long ranged, the macroscopic description of the system must include a Stokes equation that captures the effect of hydrodynamic interactions. These systems, termed ``wet'' active nematics have received much recent attention \cite{Thampi2013,Pismen2013a,Giomi2013,Schaller2013}. When the medium is frictional (such as the substrates in vibrated rods \cite{Narayan2007} or cell colonies \cite{Wu2011,Starruss2012}) or the flow induced by activity is local due to confinement (as in \cite{Pismen2013a}), the systems are termed ``dry'' active nematics and are the class of systems for which this work is relevant.

   Active nematics were first considered in the seminal work of Ramaswamy and collaborators \cite{Simha2002,Ramaswamy2003,Mishra2006,Mishra2010,Bertin2013}, who demonstrated that this system exhibits giant number fluctuations and these fluctuations render the system intrinsically phase separated. Subsequently, extensive studies have been carried out within the context of particular microscopic models, namely the ``nematic Viscek'' model \cite{Simha2002a,Chate2006,Chate2008,Ginelli2010,Peshkov2012a,Bertin2013,Ngo2013,Peruani2011,Peruani2012,Shi2013} and a system composed of reversing self-propelled rods \cite{Shi2014, Starruss2012}. These studies have delineated in detail the large scale dynamics of active nematics that are formed by the particular microscopic model.

   Our work builds on these findings by considering a minimal theory for an active nematic numerically and analytically. In particular, the equations we consider are phenomenological. Therefore, the parameters of the theory are independent of any particular microscopic model and are varied independently. We show that the curvature driven mass flux identified in \cite{Simha2002} causes the homogeneous nematic state to be unstable and leads the system to phase separate into high density and low density bands. We focus on the regime where this phenomenon is universal (independent of particular models or parameter choices), namely low energy excitations near the critical point associated with the isotropic-nematic transition. The mechanism which leads to the formation of this band structure is identified and shown to be of the same origin as those which lead to phase separation in isotropic and polar active fluids identified earlier \cite{Tailleur2008, Cates2011, Redner2013, Fily2012, Gopinath2012}.

   The layout of the paper is as follows. First, we introduce the continuum hydrodynamic theory of a generic active nematic and discuss the features that render this system inherently out of equilibrium. Then, we map out the domain of linear stability of the homogeneous nematic state and identify the mechanism that destabilizes it. Also, we report a numerical study of these hydrodynamic equations that tracks the evolution of the fully nonlinear dynamics and characterize the phase separated end state and the emergent structure of the resulting band. Finally, the primary results in this work are summarized and discussed in the context of active fluids of various symmetries.

   \section{The Continuum Theory}\label{SecEqs}

   Let us consider a collection of particles, in two spatial dimensions, that are active and interact via purely nematic aligning interactions. In the presence of a frictional medium, their microdynamics can be reliably captured by overdamped equations of motion such as in \cite{Chate2006, Sumino2012}. In the limit of length scales long compared to the particle size and time scales long compared to the interaction times, it is fruitful to consider a mean field, macroscopic description of the system, which is given by the dynamics of conserved quantities and broken symmetry variables. In the case of an overdamped active nematic of shakers, the relevant variables are the density of active units $\rho =\left\langle \sum_{\alpha}\delta \left( \mathbf{r}-\mathbf{r}_{\alpha}\right) \right\rangle $ and the nematic order parameter
   $Q_{ij}=\rho S_{ij}=\left\langle \sum_{\alpha}\left( \hat{u}_{\alpha i } \hat{u}_{\alpha j }-\frac{1}{2}\delta _{ij}\right) \delta \left( \mathbf{r}-\mathbf{r}_{\alpha}\right) \right\rangle $
   , where $\left\{ \mathbf{r} _{\alpha},\mathbf{\hat{u}}_{\alpha}\right\} $ are the positions and orientations of the shaker particles and $\left\langle {}\right\rangle $ denotes coarse-graining by averaging over microscopic lengths and times. Given the microdynamics, this coarse-graining procedure can be carried out through systematic approximations as considered in \cite{Baskaran2008, Peshkov2012a, Bertin2013, Shi2014, Starruss2012}. Alternately, one can construct the macroscopic equations based on purely symmetry considerations as in \cite{Simha2002, Ramaswamy2003}. In this work, we take the latter route, which gives us the advantage of liberating the parameters of the hydrodynamic theory from the constraints of a particular underlying microscopic model. The dynamical equations are generically of the form
   \begin{subequations}
   \begin{equation}
   \partial _{t}\rho =D\nabla ^{2}\rho +D_{Q}\bm\nabla \bm\nabla \mathbf{:}\mathbf{Q}
   \label{dtrho1}
   \end{equation}
   \begin{equation}
   \begin{array}{rcl}
   \partial _{t}Q_{ij} &=& D_{r}\left[ \alpha \left( \rho \right) -\beta \mathbf{Q}\mathbf{:}\mathbf{Q}\right]
   Q_{ij} +  D_{b}\nabla ^{2}Q_{ij}   \\
   &+&  D_{s}\partial _{k}\left( \partial _{i}Q_{kj}+\partial _{j}Q_{ik}-\delta
   _{ij}\partial _{l}Q_{kl}\right)    \\
   &+& D_{\rho }(\partial _{i}\partial _{j}-\frac{1}{2}\delta
   _{ij}\nabla ^{2})\rho
   \label{dtQ1}
   \end{array}
   \end{equation}
   \label{Eq1}
   \end{subequations}
   (where $\mathbf{A}\mathbf{:}\mathbf{B}$ refers to the contraction $A_{ij}B_{ij}$ and $\bm\nabla\bm\nabla$ is the tensor $\frac{\partial}{\partial x_i}\frac{\partial}{\partial x_j}$).

   The primary physics of the above equations is as follows. In the case of
   an equilibrium nematic, the density, a conserved quantity, has a dynamics
   generically of the form $\partial _{t}\rho =-\frac{1}{\gamma}\nabla ^{2}\frac{\delta F}{\delta \rho \left( r\right)}$, where $F$ is the free energy functional whose
   extremum is the equilibrium state and $\gamma$ is a relaxation time. This is what is referred to as Model B dynamics. By retaining every symmetry allowed term
   in the free energy, this yields a dynamics for the density given by
   \beqs
   \begin{array}{rcl}
     J_{i}=&-&D_{ij}^{0}\left( \rho ,Tr\left( Q^{2}\right) \right)\nabla _{j}\rho \\
     &-& D_{ij}^{1}\left( \rho ,Tr\left( Q^{2}\right) \right)\nabla _{j}Tr\left( Q^{2}\right). \\
   \end{array}
   \eeqs
   Here, the diffusion tensors $D^{\alpha}_{ij}$'s have the
   symmetries of the underlying nematic, i.e., $D_{ij}=A\delta _{ij}+B\left(
   \hat{n}_{i}\hat{n}_{j}-\frac{1}{2}\delta _{ij}\right) $, where $\hat{n}$ is
   the nematic director and $A$ and $B$ are potentially arbitrary functions of
   the two scalars in the theory, namely the density and the magnitude of
   orientational order $\rho S\equiv \sqrt{2TrQ^{2}}$. But, in the case of an
   active nematic, we have no extremization principle and hence the flux $J$ is
   liberated from the above constraints and hence is generically of the form
   \beqs
   \begin{array}{rcl}
     J_{i}=&-&D_{ij}^{0}\left( \rho ,Tr\left( Q^{2}\right) \right)\nabla _{j}\rho \\
     &-& D_{ij}^{1}\left( \rho ,Tr\left( Q^{2}\right) \right)\nabla _{j}Tr\left( Q^{2}\right) \\
     &-& D_{ij}^{2}\left( \rho ,Tr\left( Q^{2}\right)\right) \nabla _{k}Q_{kj}
   \end{array}
   \eeqs
   The new term here is a mass flux that arises
   because of the activity of the individual units and the resulting
   anisotropic forces produced by it. This mass flux term is non-integrable in that it cannot emerge from Model B dynamics associated with a free energy and is the central feature that
   makes an active nematic an inherently non-equilibrium system. In the present
   study we take the simplest form for the mass flux that captures the role of curvature, namely the form which was introduced by Ramaswamy et. al. \cite{Ramaswamy2003} $J_{i}=-D\nabla
   _{i}\rho -D_{Q}\nabla _{k}Q_{ki}$, where $D$ and $D_{Q}$ are constants
   independent of the dynamical fields. This results in Eq. (\ref{dtrho1}) for the
   dynamics of the density field.

   The nematic order parameter $Q_{ij}$ has a dynamics analogous to an
   equilibrium nematic, i.e., $\partial _{t}Q_{ij} =-\frac{1}{\gamma}\frac{\delta F}{\delta Q_{ij} \left( r\right)}$, where $F$ is a Ginzburg-Landau free energy functional as above. The first terms on the right hand side of Eq. (\ref{dtQ1}) give
   rise to a second order phase transition from a disordered isotropic state to
   an ordered nematic state when $\alpha \left( \rho \right) $ changes sign. $%
   D_{b}$ and $D_{s}$ are related to the Frank elastic constants associated
   with bend and splay deformations. The term proportional to $D_{\rho }$ is a kinetic term arising due to the inherent anisotropy in diffusive processes in a nematic.  The existence of this term implies that when $Q=0$, $(\partial_i\partial_j-\frac{1}{2}\delta_{ij}\partial^2)\rho=0$, and so the density must become homogeneous. Any inhomogeneous state will have local nematic order even though the state is globally isotropic.  An alternative description, where this term vanishes in the isotropic limit, is discussed briefly in Appendix \ref{iso_limit}.

   In the following, we non-dimensionalize our equations by picking the units
   of time to be given by $1/D_{r}$, the rotational diffusion time and our
   length scale to be $\sqrt{D/D_{r}}$, a diffusion length. Also, we choose $%
   \alpha \left( \rho \right) =\left( \rho -1 \right) $ and $\beta \left( \rho
   \right) =\frac{1}{\rho ^{2}}\left( 1+\rho \right) $, thereby setting the
   density at which the isotropic nematic transition occurs to $1$ and ensuring
   that when $\rho \gg 1$ the nematic order saturates to a finite value. Further, to minimize
   the number of independent parameters in our study, we make a one elastic constant approximation,
   namely $D_{s}=D_{b}\equiv D_{E}$. The dimensionless dynamical equations then become
   \begin{subequations}
   \begin{equation}
   \partial _{t}\rho =\nabla ^{2}\rho +D_{Q}\bm\nabla \bm\nabla \mathbf{:}\mathbf{Q}
   \label{dtrho2}
   \end{equation}
   \begin{equation}
   \begin{array}{rcl}
   \partial _{t}Q_{ij} &=& \left[ \alpha -\beta \mathbf{Q}\mathbf{:}\mathbf{Q} \right] Q_{ij} + D_{E}\nabla ^{2}Q_{ij} \\
   &+&  D_{E}\partial _{k}\left( \partial _{i}Q_{kj}+\partial _{j}Q_{ik}-\delta_{ij}\partial _{l}Q_{kl}\right)  \\
   &+&  D_{\rho}(\partial _{i}\partial _{j}-\frac{1}{2}\delta _{ij}\nabla^{2})\rho
   \end{array}
   \label{dtQ2}
   \end{equation}
   \label{fullsimp1}
   \end{subequations}
   These simplified equations are studied analytically and numerically in the sections below.

   \section{Linear Stability Analysis}\label{lsa}

   The dynamical equations Eq. (\ref{fullsimp1}) admit two homogeneous steady states, an
   isotropic state when $\rho <1$ and a uni-axial nematic state when $\rho \geq 1
   $. Let us focus on the high density ordered state and without loss of
   generality let us consider a nematic state where the direction of broken
   symmetry is along the $x$ axis of our coordinate system. Then,
   small fluctuations about this homogeneous steady state can be parametrized
   as $\rho =\rho _{0}+\delta \rho \left( \mathbf{r},t\right) $, $Q_{xx}=\frac{1%
   }{2}\rho_0S_{0}+\delta Q_{xx}\left( \mathbf{r},t\right) $ and $Q_{xy}=0+\delta
   Q_{xy}\left( \mathbf{r},t\right) $ where $S_{0}=\sqrt{\frac{2(\rho
   _{0}-1)}{\rho _{0}+1}}$. Introducing a spatial Fourier transform $\widetilde{%
   X}\left( \mathbf{k},t\right) =\int d\mathbf{r}e^{i\mathbf{k}\cdot \mathbf{r}%
   }X\left( \mathbf{r},t\right) $, the linearized dynamics of fluctuations in
   density and order parameter takes the form
   % Widetext format for a full-page equation in pre
   \begin{widetext}
   \begin{equation}
   \partial _{t}\left[%
   \begin{array}{c}
   \delta \tilde{\rho} \\
   \delta \tilde{Q}_{xx} \\
   \delta \tilde{Q}_{xy}%
   \end{array}%
   \right]=-\left[%
   \begin{array}{ccc}
   k^{2} & D_{Q}k^{2}\cos (2\phi ) & D_{Q}k^{2}\sin (2\phi ) \\
   \frac{1}{2}D_{\rho}k^{2}\cos (2\phi )-C_{0} & 2D_{E}k^{2}+2\alpha _{0} & 0 \\
   \frac{1}{2}D_{\rho}k^{2}\sin (2\phi ) & 0 & 2D_{E}k^{2}%
   \end{array}%
   \right]\left[%
   \begin{array}{c}
   \delta \tilde{\rho} \\
   \delta \tilde{Q}_{xx} \\
   \delta \tilde{Q}_{xy}%
   \end{array}%
   \right]  \label{lsa_array}
   \end{equation}%
   \end{widetext}
   where $\phi $ is the angle between the director and the spatial gradient
   vector $\mathbf{k}$, $\alpha _{0}=(\rho_0-1)$ and $C_{0}=\sqrt{\frac{2(\rho_{0}-1)}{\rho_{0}+1}}\Big(\frac{\rho_{0}^{2}+\rho_{0}-1}{\rho_{0}+1}\Big)$.

   \begin{figure}
     \centering
     \includegraphics[width=0.8\columnwidth]{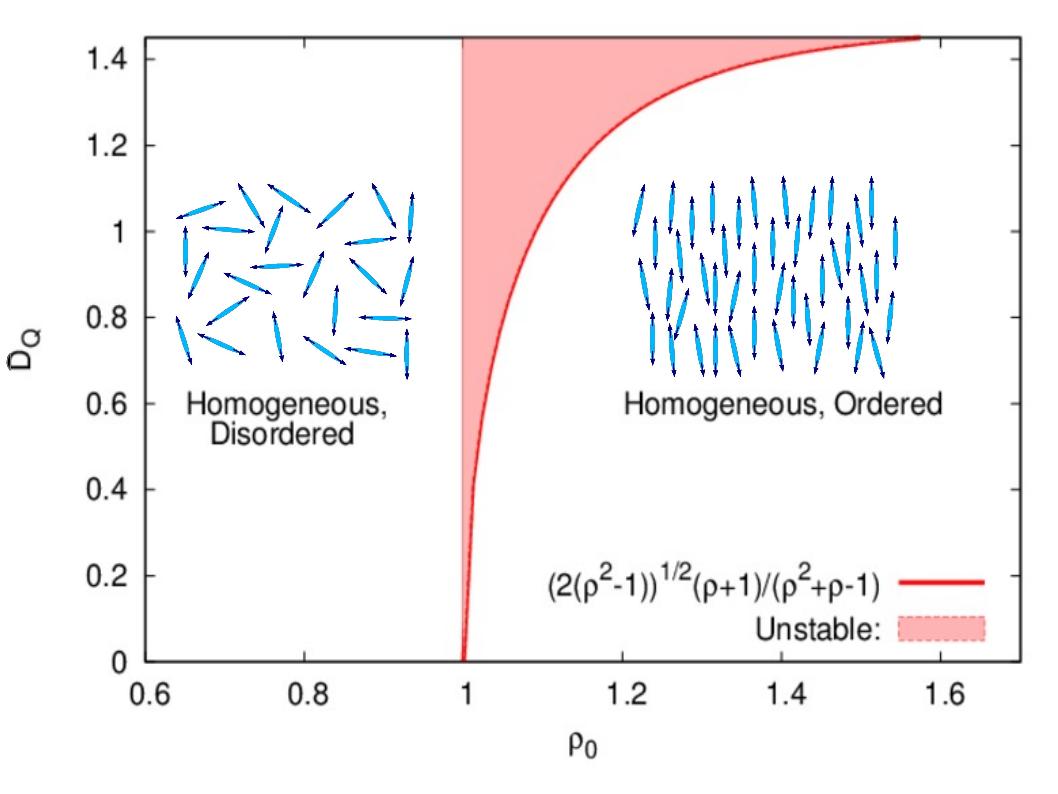}
     \caption{(color online) Phase behavior of the system as a function of the mean density of the system ($\rho_0$) and the activity ($D_Q$).  Below the critical density for the order disorder transition (i.e., $\rho<1$) the homogeneous disordered state is stable.  For any density $\rho\geq1$, there is an activity $D_Q$ above which the homogeneous ordered state is unstable.  This region, in which there is no stable homogeneous state, is shown in red.}
     \label{stability}
   \end{figure}

   While we can readily analyze the cubic characteristic equation of this
   linear system (see Appendix \ref{MaxEval}), the physics at play is best exposed by considering spatial
   fluctuations in a direction orthogonal to the mean nematic director, i.e.,
   consider $\phi =90^{\circ }$. In this case the director fluctuations $\delta
   \tilde{Q}_{xy}$ decouple and are always diffusive and stable. The relevant
   dynamics of the system is now in the $\left( \rho ,Q_{xx}\right) $ subspace.
   The eigenvalues associated with density and magnitude of order fluctuations
   take the form
   % Widetext format for a full-page equation in pre
   \begin{widetext}
   \begin{align*}
     \lambda_{\pm}=-\frac{1}{2}\Bigg(&2\alpha_{0}+(2D_{E}+1)k^{2} \\
	     &\pm\sqrt{ \big(2\alpha_{0}+(2D_{E}+1)k^{2}\big)^{2} -4k^{2}\big(2\alpha_{0}-D_{Q}C_{0}+ k^{2}(-\frac{1}{2}D_{\rho}D_{Q}+2D_{E})\big) }\Bigg)
   \end{align*}
   \end{widetext}
   Clearly $\lambda _{+}$ is always negative and the associated fluctuations decay to homogeneity. $\lambda _{-}$ on the other hand, in the long wavelength limit, takes the form $\lambda_{-}\sim-\frac{1}{2}k^{2}\big(-D_{Q}\frac{C_{0}}{\alpha_{0}}+2\big)+\mathcal{O}(k^{4})$
   and hence becomes positive whenever
   ${  D_{Q}>\frac{2\alpha _{0}}{C_{0}}=\Big(2(\rho _{0}^{2}-1)\Big)^{\frac{1}{2}}\Big(\frac{\rho _{0}+1}{\rho_{0}^{2}+\rho _{0}-1}\Big)  }$
    Fig. \ref{stability} shows a plot of this threshold value of
   activity as a function of the mean density of the system. Note that the
   threshold goes to zero as the order disorder transition is approached and
   hence in the vicinity of the critical point, arbitrarily small values of the
   activity destabilize the homogeneous nematic. The vanishing of the threshold for the onset of this instability is independent of the detailed form of $\alpha(\rho)$ and is a universal feature of the dynamics of active nematics.

   The physics of this instability of the homogeneous nematic state can be
   understood as follows. The order parameter Eq. (\ref{dtQ2}) has a second order mean
   field transition from an isotropic to a nematic state that is controlled by
   the density of the system. This density, in the context of conventional
   nematics is governed by Model B dynamics generated from a free energy functional. Such dynamics always causes fluctuations in the density to decay rapidly to homogeneity. Hence, the mean density effectively acts as an external control parameter that determines the magnitude of order in the system. But, in the case of active
   nematics,  the dynamics of the density is coupled non-trivially to the magnitude and direction of order in the system through the non-equilibrium curvature induced mass flux term
   proportional to $\nabla _{j}Q_{ij}$,  arising due to the presence of the microscopic active forces that drive the system. It causes a feedback between fluctuations in order and density such that homogeneous states are destabilized. As we will show below, a consequence of this anomalous coupling is that the local density and the amount of order in the systems are now both controlled by the strength of the activity $D_Q$.

   The discussion up to this point focused on spatial fluctuations orthogonal to the direction of nematic order. We can of course analyze the eigenmodes for arbitrary directions of spatial fluctuations and we do so
   perturbatively in the wavevector in Appendix \ref{MaxEval}. We find that a range of wavevectors in a sector $[\phi_{min},90^{\circ}]$ with respect to the nematic director go unstable, depending on the specific values of the different parameters. Fluctuations perpendicular to the director (i.e., $\phi=90^{\circ}$) are, however, dominant for most parameters and are the only relevant dynamics close to the critical density. In the remainder of this presentation we
   explore Eq. (\ref{fullsimp1}) numerically in the unstable regime and
   characterize the resulting inhomogeneous end states.

   \section{Emergent Structures - Phase-separated Bands}
   \label{NumPhen}

   In order to understand the consequences of the linear instability discussed above
   on the formation of emergent structures in an active nematic fluid, we
   solved Eq. (\ref{fullsimp1}) numerically. The integration was performed using an Euler method,
   forward time centered space (FTCS) scheme on a grid with periodic boundaries. Typical values
   used for the time step and spatial resolution were 0.01 diffusion times and 0.4
   diffusion lengths. The system size ranged from 200 to 1000 diffusion lengths on a side. To further reduce the number of parameters and to best illustrate the principal mechanisms at play, we first report the results for the case $D_\rho$, and $D_E$ are fixed at $1.0$. The consequence of varying $D_\rho$ and $D_E$ is discussed in Appendix \ref{GenNA}.

   \begin{figure}[h!]
     \begin{center}
       \subfigure[]
           {\includegraphics[width=0.52\columnwidth]{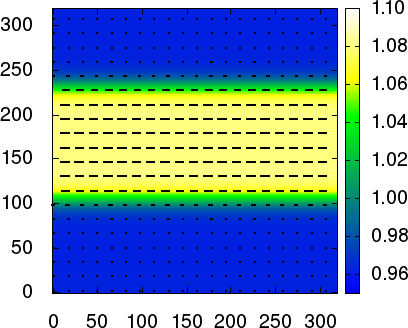}}
       \subfigure[]
           {\includegraphics[width=0.45\columnwidth]{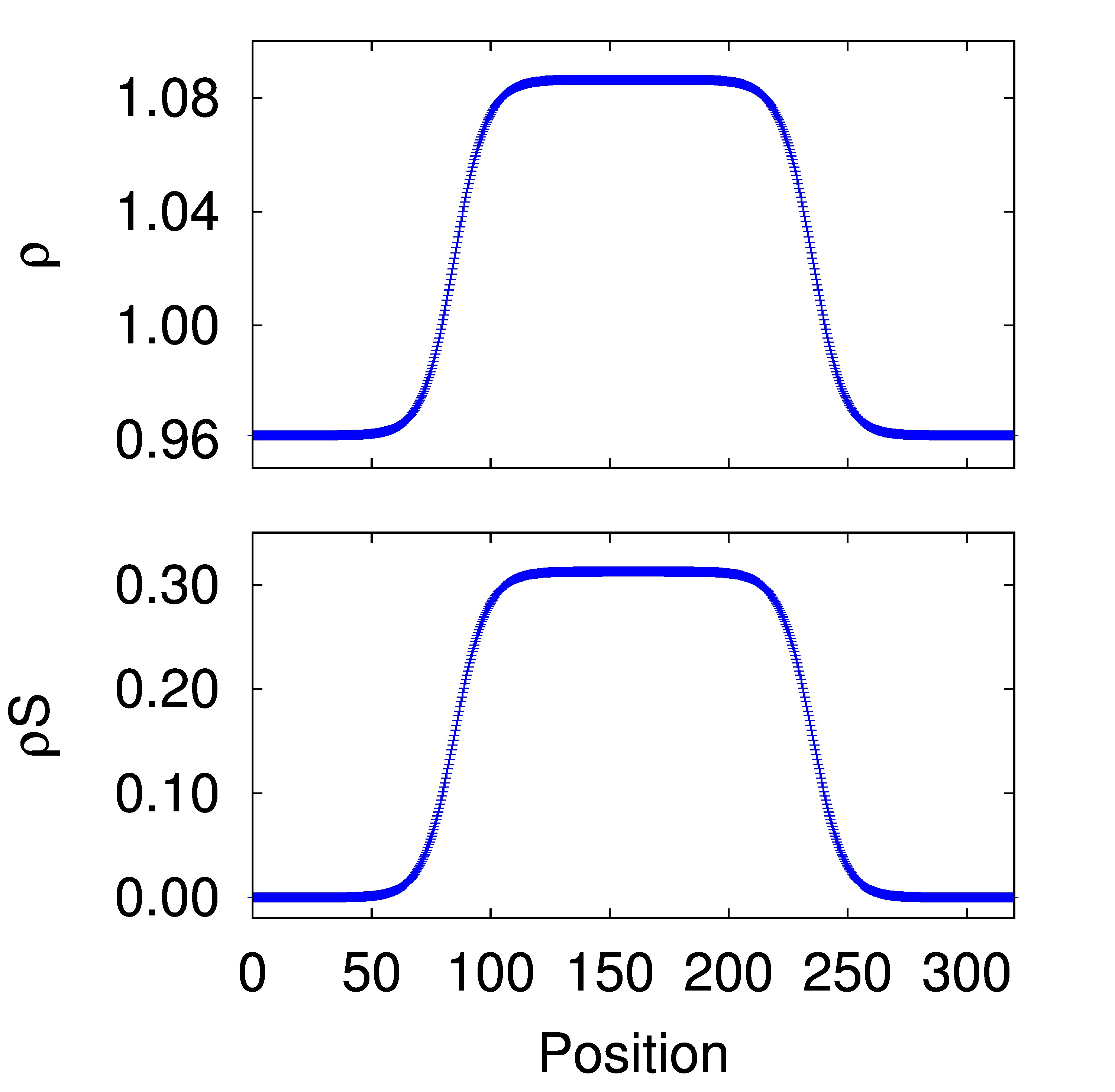}}
     \end{center}
     \caption{(color online)(a) A plot of the density and order of a typical system ($\rho_0=1.01$, $D_Q=0.8$) that has phase-separated into bands. The lines show the magnitude (by the length) and direction of nematic ordering.  The light region is a band of high density with nematic ordering along the band. The axes show the position in the system in dimensionless `diffusion lengths' and the scale bar shows the density. (b) Profiles of the density (top) and order ($\rho S$, on bottom), taken perpendicular to the direction of ordering in the bands.
     }
     \label{phasesep}
   \end{figure}

   The primary finding of our numerical analysis is as follows:  The typical inhomogeneous state we find when the activity $D_{Q}$ is greater than the threshold for onset of the instability is shown in Fig. \ref{phasesep}. The system develops inhomogeneous bands of alternating low and high density.  These bands coarsen, and the steady state which the system typically reaches is one band which spans the system.  The only characteristic size associated with these bands is the length scale associated with the interface between high and low density, which can be seen in the band profile in Fig. \ref{phasesep}.

One way to understand these structures is to consider the reduced set of equations that describe the dynamics of the density and the magnitude of ordering namely

   \begin{subequations}
   \beqs
   \partial _{t}\rho = \big(\partial_x^2+\partial_y^2\big) \rho
   +\frac{1}{2}D_Q \big( \partial_x^2-\partial_y^2 \big) \rho S
   \eeqs
   \beqs
   \begin{array}{rcl}
   \partial_{t}\rho S &=& \big( \alpha(\rho) -\beta(\rho)\rho^2S^2) \rho S+2D_\rho\big( \partial_x^2 - \partial_y^2 \big) \rho  \\
   &+& 2D_E\big( \partial_x^2 + \partial _y^2 \big) \rho S
   \end{array}
   \eeqs
   \end{subequations}
    where we have assumed that the mean nematic order lies along the $x$ axis of our coordinate system. These reduced equations admit a stationary solution homogeneous in $x$, and having a profile of high density, high order region embedded between low density isotropic regions along the $y$ direction as has been shown in \cite{Bertin2013}. Here we present a complementary analysis.

\begin{figure}[h!]
     \begin{center}
       \subfigure[]{ \includegraphics[width=0.8\columnwidth]{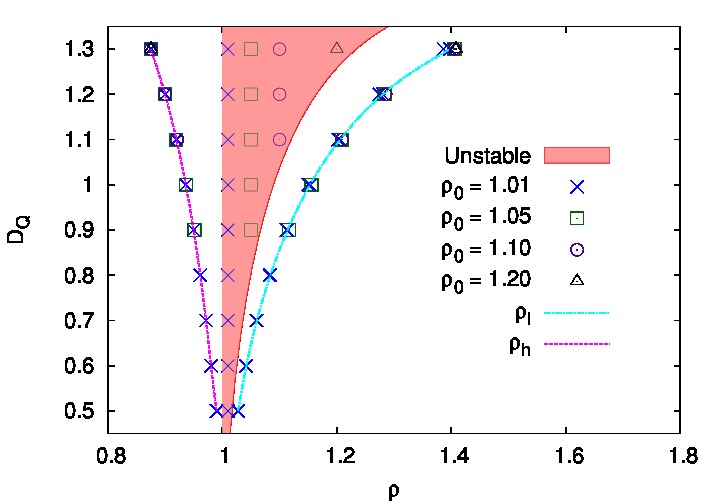} \label{selfreg}}
       \subfigure[]{
         \includegraphics[width=0.8\columnwidth]{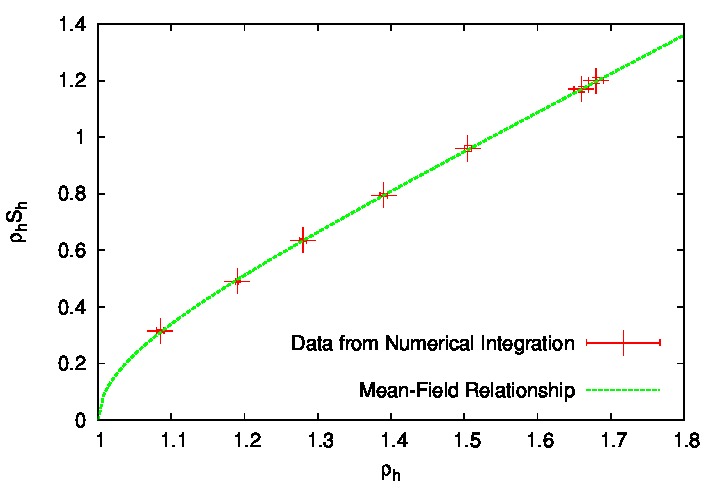} \label{meanfield}}
     \end{center}
     \caption{(color online) (a) For fixed activity, several initial densities in the unstable region are chosen. The final densities in the banded state ($\rho_h$ and $\rho_l$) are found to be independent of the initial density, as the points for different initial densities fall on the same curve.  (b) The measured value of order ($\rho S$) as a function of $\rho_h$. The solid line is the mean field prediction $\rho _{h}\sqrt{\frac{2(\rho _{h}-1)}{\rho _{h}+1}}$. Excellent agreement is found substantiating the picture that the bands are just phase coexistence between a high density nematic and a low density isotropic state.
     }
     \label{heatmap}
   \end{figure}

   First note that as shown in Fig. \ref{selfreg}, the density of the nematic band is independent of the value of the homogeneous density of the system but instead is determined entirely by the strength of the activity $D_{Q}$. Secondly note that as shown in Fig. \ref{meanfield}, we measure the magnitude of nematic order in the bands and find that $(\rho S)_{band}=\rho _{h}\sqrt{\frac{2(\rho _{h}-1)}{\rho _{h}+1}}$, i.e., $S$ is related to the density by the same mean field relation in the homogeneous theory, but now the mean density is replaced by $\rho _{h}$, the density in the band. Note that $\rho _{h}$ is such that the nematic state is no longer unstable to the linear instability at this value of the activity. This suggests that the formation of this banded structure can be viewed as the system phase separating into a high density nematic state and a low density isotropic state both of which are stable, reminiscent of gas-liquid coexistence. The amount of order in the system now is determined by the strength of the activity $D_Q$ (through $\rho_h$) and not by the mean density.

   This simple picture holds for a range of activities when the density is close to the critical density $\rho_c$, for our choice of parameters that set all the spatial relaxation mechanisms equal to each other (i.e., $D_\rho=D_E=D=1$). However, for larger values of $D_\rho/D_E$ these bands can become unstable, leading to complex dynamical structures.  These structures are similar to those which have been discussed in the context of reversing rods by Shi et. al. \cite{Shi2014} and the nematic Vicsek model by Ngo et. al. \cite{Ngo2013} and are discussed briefly in Appendix \ref{GenNA}.

   \section{Universality of activity-induced phase separation}

   We have shown that the homogeneous ordered state of an active nematic is unconditionally unstable near the critical density $\rho_c$ for the onset of order. This instability arises because of the dynamic coupling between the density and order through the curvature driven mass flux. Near the critical density, the system phase separates into a high density nematic and a low density isotropic fluid with the direction of order being perpendicular to the direction of the density gradient. The density and amount of order in the nematic phase is determined by the magnitude of activity.

   This kind of phase separation phenomena are ubiquitous in model systems of active fluids. In active polar fluids, composed of self-propelled particles with polar aligning interactions, theory and simulations show that the system phase separates into high density and low density stripes with the high density regions forming propagating solitary waves \cite{Gregoire2004,Chate2008,Bertin2009,Solon2013,Gopinath2012}. These propagating waves have been observed in model experimental systems \cite{Schaller2010, Saragosti2011, bricard2013emergence}. In the case of active isotropic fluids composed of self propelled spheres with no aligning interactions, athermal phase separation into a dense phase and a dilute phase has been extensively reported and discussed \cite{Fily2012,
   Redner2013, Tailleur2008, Cates2010, Stenhammar2013}. This phenomenon has been observed in experiments of diffusophoretic Janus colloids as well \cite{Buttinoni2013}. The underlying physical mechanism that leads to this phase separation is the presence of a self-replenishing velocity along one direction of the body axis of a particle and the consequent persistent collisions that result among such active particles \cite{Redner2013}.

   From a completely macroscopic point of view, however, the phenomenon of phase separation in active fluids of different symmetries can be unified as follows. Within a Ginzburg Landau approach, an equilibrium system near a critical point is described by a free energy that is a functional of an order parameter. The nature of the equilibrium state is determined by the value of a control parameter say $\alpha$. When $\alpha\leq \alpha_c$, the free energy is a minimum when the order parameter is zero, while when $\alpha\geq \alpha_c$ the minimum and hence the equilibrium state is one where the order parameter is finite. The control parameter in these theories is a density or a temperature whose dynamics is relaxational  (model B). Therefore the control parameter rapidly relaxes to homogeneity, and hence can effectively be tuned externally. What happens in the case of active fluids is that the relaxational Model B dynamics of the control parameter is now modified by the activity. Anomalous non integrable terms in its dynamics give rise to non-trivial coupling to the order parameter (magnitude and direction) of the system. Let us explicate this statement by considering specific examples.

In the active nematic system studied in this work, the order parameter is the nematic ordering tensor $Q_{ij}$. As mentioned above, the dynamical equation associated with this order parameter (to the order considered here) is indeed the same as that for an equilibrium nematic and can be schematically written as $\partial_t Q_{ij}=-\frac{1}{\gamma}\frac{\delta F[\rho,Q_{ij}]}{\delta Q_{ij}}$. The key active feature of the dynamics of this fluid is that the density is now anomalously coupled to $Q_{ij}$ through the curvature induced mass flux. This inherently non-integrable term is the reason the active nematic phase separates into a low density isotropic phase and a high density nematic phase whose properties are now controlled by the strength of the activity.

In the context of a polar active fluid, the order parameter is a vector $\mathbf{P}$ that measures orientational ordering in the velocity field of the self-propelled entities \cite{Vicsek1995, Gopinath2012}. The dynamics of this system can generically be written as $\partial_t \mathbf{P} = -\frac{1}{\gamma}\frac{\delta F[\rho,\mathbf{P}]}{\delta \mathbf{P}}+N[\rho,\mathbf{P}]$, i.e., a part composed of exactly what we would have in the equilibrium case and a new truly nonequilibrium piece $N$. As has been shown in \cite{Gopinath2012,Solon2013}, the nonequilibrium piece is irrelevant to the observed phase separation behavior, which arises again due to the fact that the dynamics of the density is anomalously coupled to the magnitude of the order parameter through a non integrable term of the form $\partial_{t} \rho = -v\nabla\cdot \rho \mathbf{P}$. This coupling causes the system to phase separate into a low density isotropic phase and a high density polar phase whose properties are controlled by the strength of the activity.  %tsTt

Finally in the case of the isotropic fluid, the order parameter is a density while the control parameter is a chemical potential (as in a gas-liquid system). As has been shown in the works of \cite{Tailleur2008, Cates2010, Stenhammar2013}, the self-propulsion of the active particles serves as a chemical potential in the case of an active isotropic fluid. But, the local density determines the local self-propulsion speed and therefore the effective control parameter, and hence the analogy of the control parameter being rendered dynamically coupled to the order parameter extends to this class of active fluids as well.

The consequence of the above feature is that even though the dynamics of the order parameter in the system is the relaxational dynamics familiar from the near equilibrium context, due to the anomalous dynamics of the control parameter, the state of the system is controlled by the strength of the activity.  Hence, as is the case for active nematics \cite{Mishra2006}, active fluids are intrinsically phase separated. The structure of the end state is determined by the symmetry of the interactions among the fluid particles, with spherical droplets in the case of the isotropic active fluid, ordering orthogonal to the interface in the case of polar fluids and ordering along the direction of the interface for nematic fluids.

\section{Summary}

In this work we consider a minimal model for an active nematic fluid and show that the fluid undergoes phase separation beyond a given threshold for activity and that the threshold vanishes as the critical density for the order disorder transition is approached. In the coexistence region, we show that the system forms bands of nematic ordered regions where the ordering is orthogonal to the direction of the density gradient, i.e., the nematic is oriented parallel to the interface. We identify the macroscopic mechanism for this phase separation and relate it to those found in the context of polar and isotropic active fluids.

% is only one appendix.
%\appendix
%\section{}

\appendix
\section{Generalized Linear Stability Analysis}\label{MaxEval}

In Section \ref{lsa} we considered the stability of the homogeneous, ordered state with respect to spatial fluctuations that were orthogonal to the direction of nematic ordering (i.e., the angle between the spatial gradient vector ($\mathbf{k}$) and the nematic director ($\hat{n}$) fixed at $\phi=90^\circ$). In the following we consider arbitrary spatial fluctuations and show that the lowest threshold value for $D_Q$ and the direction of the fastest growing wavevector lie at $\phi=90^\circ$, when the density is close to the critical value. Far away from the critical density, there is a dependence on the particular parameters which are characterized below as well.

The dynamics of arbitrary spatial fluctuations about the homogeneous ordered state is characterized by the linear system of equations Eq. (\ref{lsa_array}). As we are interested in the dynamics on the longest length scales, let us consider roots of the cubic characteristic equation for the linear system in the long wavelength limit. The perturbative solution to order $k^2$ is found to be
\begin{widetext}
\begin{align*}
\lambda_{\pm}=-\frac{k^2}{4\alpha_0}\Bigg(&C_0 \cos(2\phi)D_Q+2\alpha_0(1+2D_E) \\
& \pm\sqrt{\big(C_0\cos(2\phi)D_Q+2\alpha_0(1-2D_E)\big)^2+8\alpha_0^2\sin^2(2\phi)D_QD_\rho}\Bigg)+\mathcal{O}(k^4)
\end{align*}
and
\[
\lambda_3=-2\alpha_0+k^2\Big(-2D_E+\frac{C_0}{2\alpha_0}D_Q\cos(2\phi)\Big)+\mathcal{O}(k^4)
\]
\end{widetext}
We can readily establish that $\lambda_-$ is the only eigenvalue that changes sign in the small wavevector limit and thereby leads to an instability. The threshold for the onset of this instability is identified as
\beq
D_Q>\frac{4\alpha_0D_E}{\alpha_0\sin^2(2\phi)D_\rho-2C_0\cos(2\phi)D_E}.
\label{dqeq}
\eeq
\begin{figure}
  \centering
  \includegraphics[width=\columnwidth]{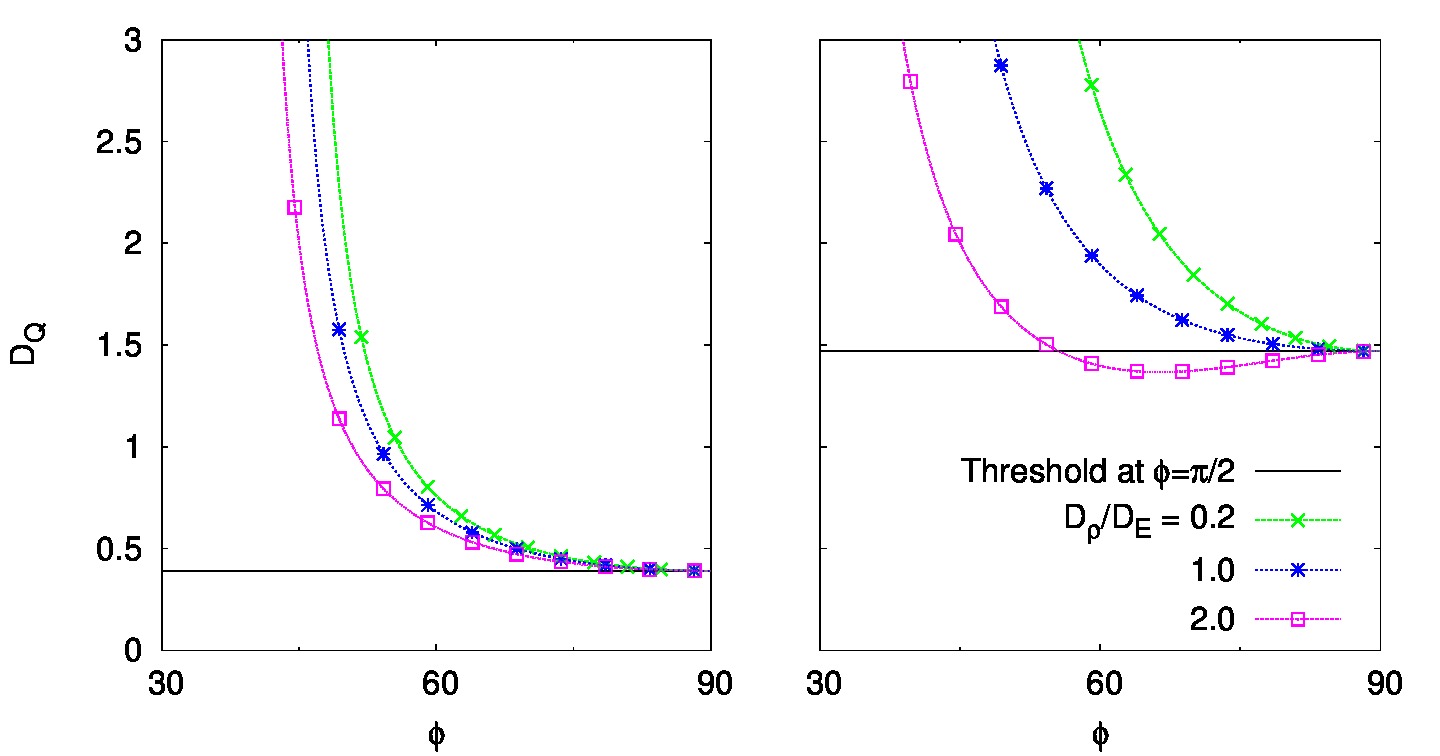}
  \caption{(color online) Plot of $D_Q(\phi)$, the value of activity at which spatial fluctuations in the direction $\phi$ become unstable. Left panel : $\rho=1.01$, close to the critical point, when $D_Q$ is monotonic and is smallest in the $\phi=90^\circ$ direction. Right panel : $\rho=2.0$, $D_Q$ becomes non monotonic when the Frank elastic constant becomes small compared to the kinetic terms. The horizontal line at $D_Q(90^\circ)$ is a guide to the eye.}
  \label{AppendixFig1}
\end{figure}
Thus, for a fixed activity $D_Q$ a range of wavevectors become unstable.

The content of the above threshold condition can be explicated as follows. %spatial fluctuations in different directions go unstable at different values of activity $D_Q$.
First let us fix $\phi$,
the direction of the spatial fluctuation and characterize $D_Q(\phi)$, the value of activity at which spatial fluctuations in this direction go unstable. This is plotted in Fig. (\ref{AppendixFig1}). Whenever $g_0\equiv\frac{D_{\rho}}{2D_{E}}\frac{\alpha_0}{C_0}<1$, which is true for densities close to the critical density for the onset of order, $D_Q(\phi$) is a monotonic function with a minimum at $\phi=90^\circ$, i.e., spatial fluctuations orthogonal to the direction of ordering go unstable at the lowest value of activity. On the other hand, when $g_0\geq 1$, the spatial fluctuation that goes unstable first as we ramp up activity from zero is now $\phi\sim \frac{1}{2} \arccos(1/g_0)$. This happens far from the critical density at which point we should expect that universality is no longer applicable and the dynamics becomes dependent on the details of the parameters of the system.

\begin{figure}
  \centering
  \includegraphics[width=\columnwidth]{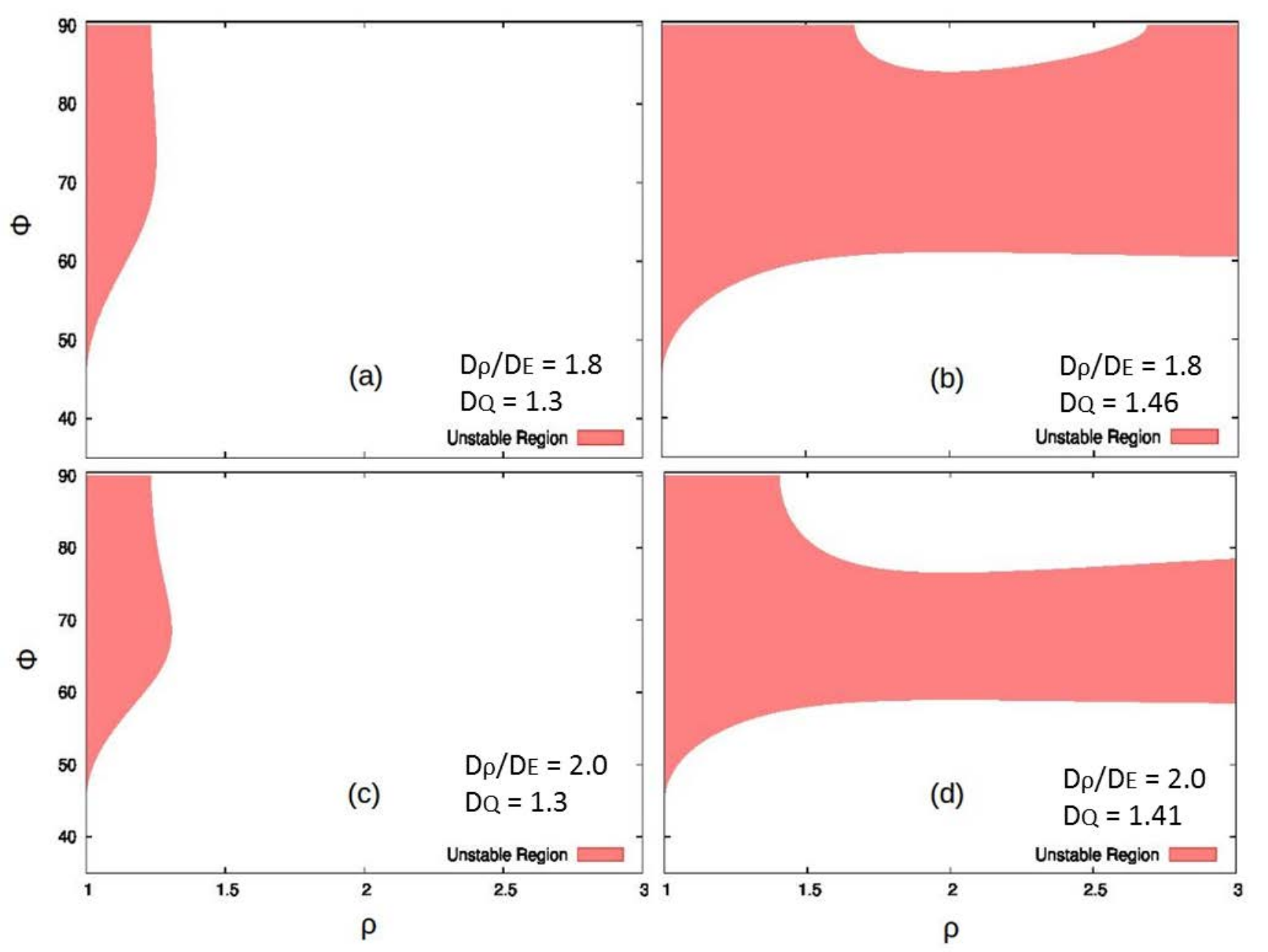}
  \caption{(color online) Identifying the directions in space associated with destabilizing fluctuations for different choices of parameters of the continuum theory. When $D_{\rho} /D_E$ becomes large, the instability persists to large values of density and is spread over a wider range of $\phi$}
  \label{AppendixFig2}
\end{figure}

Another fruitful representation of the instability condition (Eq. \ref{dqeq}) is to consider fixed values of parameters $D_Q$, $D_{\rho}$ and $D_E$, and identify the range of wavevector directions $\phi$ that are unstable to fluctuations at any given value of mean density $\rho$. This is illustrated in Fig. \ref{AppendixFig2}. The range of spatial directions associated with fluctuations that destabilize the system is determined by the ratio $g_0$ defined earlier. For $g_0>1$, the wavevectors close to $\phi=90^\circ$ are unstable. When we move to smaller values of $g_0$ wavevectors in a much wider sector of spatial directions become unstable. As the activity $D_Q$ becomes large, the instability extends to higher densities far from the critical density as well.

Even though a large number of modes go unstable depending on the details of the parameters, the physics of the system in the unstable region will be controlled by the fastest growing wavevector. We find that the fastest growing mode is along the $\phi=90^{\circ}$ whenever $D_E+\frac{C_0}{4}D_Q \geq \alpha_0 \big( \frac{\alpha_0}{C_0}D_\rho+\frac{1}{2} \big)$. This relationship is always satisfied for densities close to the critical density. Far from the critical density, there exists values of parameters for which the fastest growing mode shifts to
 \[ \cos(2\phi)=-\frac{\xi}{\gamma}\Big[ 1+\sqrt{1+\frac{\gamma}{\xi^2}} \Big] \] where $\gamma=8\frac{D_\rho}{D_Q}(\frac{\alpha_0}{C_0})^2-1 $ and $\xi=\frac{4D_E-2\alpha_0}{C_oD_Q}$.  Since the body of the work focuses on the universal regime close to the critical density and focuses on low energy excitations for which our quadratic in gradients theory is convergent and well behaved, this shift in the fastest growing wavevector and its consequence to the emergent structure is not probed.

\section{Generalized Numerical Analysis: Role of specific parameters}\label{GenNA}

\begin{figure}[h]
  \centering
  \vspace{0.60cm}
  \includegraphics[width=\columnwidth]{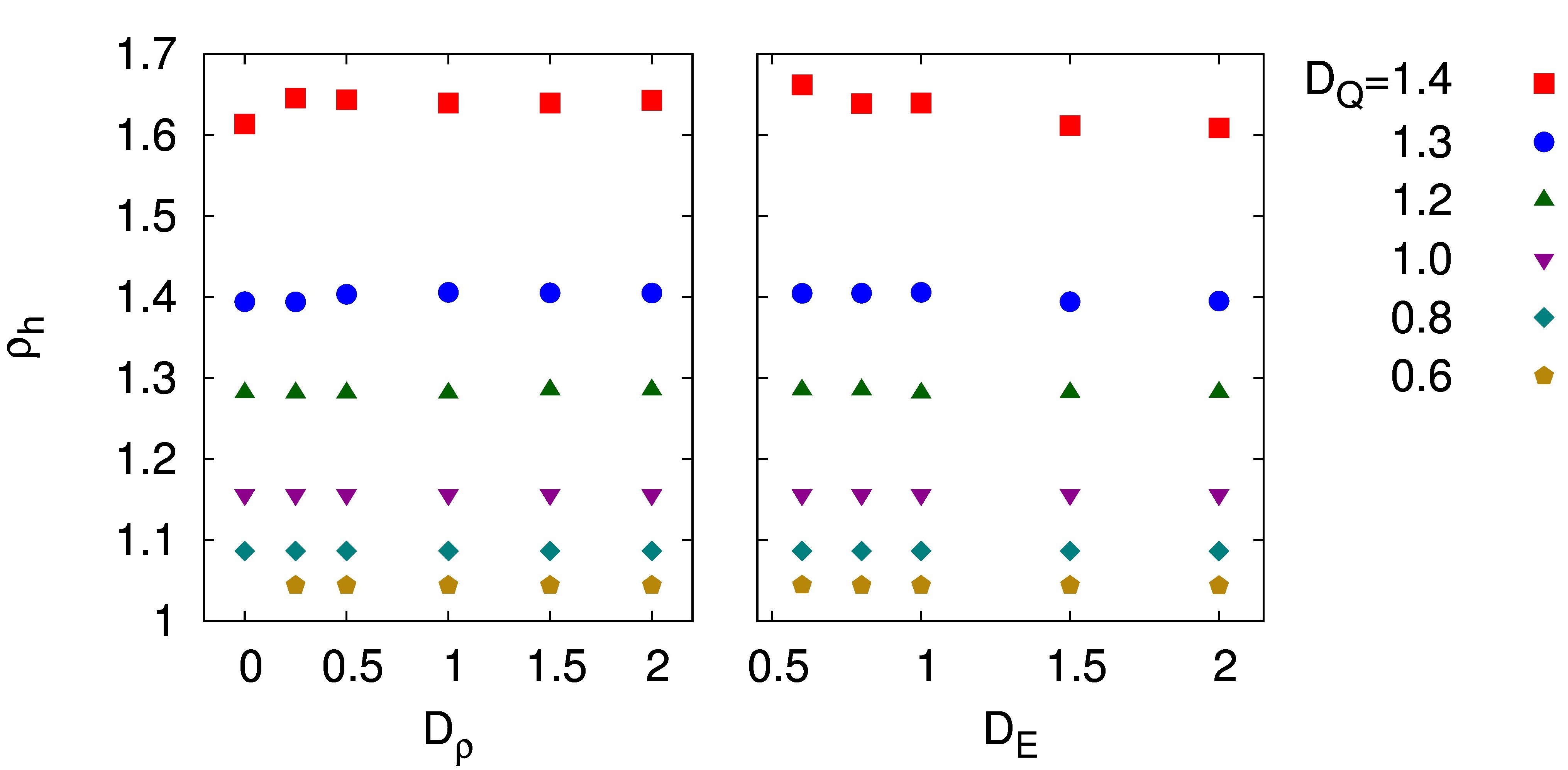}
  \caption{(color online) The density of the ordered phase ($\rho_h$) is shown for a range of $D_\rho$, $D_E$, and $D_Q$ respectively. $\rho_h$ is insensitive to $D_\rho$ and $D_E$, especially for small $D_Q$.
  }
  \label{Insensitive}
\end{figure}

The phase separation and structure formation which is discussed in the body of this paper depends entirely on $D_Q$ for low energy excitations (small $D_Q$) near the critical point.  The density contrast ($\rho_h-\rho_l$) is insensitive to other parameters, especially for small $D_Q$, as is shown for the density of the ordered phase ($\rho_h$) in Fig. \ref{Insensitive}.  The density of the disordered phase ($\rho_l$) is similarly insensitive to these other parameters, with the values of $\rho_l$ showing less deviation than those of $\rho_h$.

  %The deviation of $\rho_h$ from a constant value for larger $D_Q$ (ie $D_Q=1.4$ and $1.3$) may stem from the fact that the density does not increase smoothly to a plateau for these values of $D_Q$ as it does in Fig. \ref{phasesep}(b), but overshoots and returns to the plateau.

\begin{figure}[h]
  \centering
   \vspace{0.60cm}
   \includegraphics[width=\columnwidth]{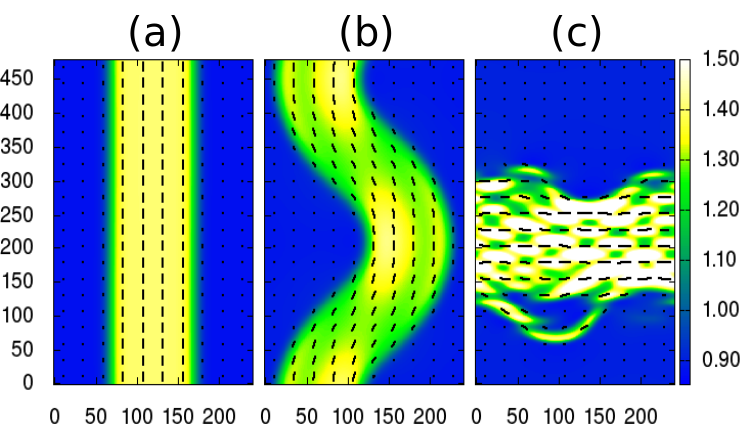}
  \caption{(color online) The plots above show the progression of the structures which form as $D\rho/D_E$ is increased.   The density is represented as a heat map, and the magnitude and direction of order is represented by the length and orientation of the lines (as in Fig. \ref{phasesep}) after $90,000$ diffusion times. These systems all have parameters $\rho_0=1.10$, $D_Q=1.30$ and $D_E=1.20$.
(a) For $D_\rho=0.80D_E $, the band of the ordered phase is stable. (b) For $D_\rho=1.20D_E$ the band is unstable to a large wave-length instability which causes it to bend and eventually break. (c) For $D_\rho=2.50D_E$ the band breaks down quickly and a structure with fluctuations on a much smaller length scale forms.  This structure is dynamical, and the order at the edges fluctuates, but it persists for over a hundred thousand of diffusion times.
}
  \label{StabilityImage}
\end{figure}

 Away from the regime of small $D_Q$ and $\rho_0\sim\rho_c$, the stability of the band structure depends on values of $D_E$ and $D_\rho$. When the system samples densities which are further from the critical density, as it does for larger values of $D_Q$, the band structure becomes unstable, and complex dynamical structures can form (see Fig. \ref{StabilityImage}).  These structures are similar to the ones seen for the reversing rod model discussed by Shi et. al. and the nematic Vicsek model considered in Ngo et. al. \cite{Shi2014,Ngo2013}. For the range of Frank elasticities which were used ($0.6 \leq D_E \leq 2.0$) the stability of the bands depended only $D_\rho/D_E$ rather than both parameters independently. In order to examine the phenomenology as a function of $D_\rho/D_E$, the system was initialized in a band with the order along the long dimension of the system (as it is in Fig \ref{StabilityImage}(a)), and a Gaussian profile of density along the shorter dimension.  The average density was chosen to be slightly less than halfway between $\rho_l$ and $\rho_h$ for each value of $D_Q$ so as to have an aspect ratio of about 4 for the band.

The progression of structures which are seen as $D_\rho/D_E$ is increased can be described as follows.  When this ratio is small the band structure is stable and the density and order profiles relax to a steady state like the ones seen in Fig. \ref{StabilityImage}(a) and \ref{phasesep}(b).  As this ratio is increased the band becomes unstable to a long-wavelength fluctuation that causes the band to bend, as seen in Fig. \ref{StabilityImage}(b), then break.  These bands, which bend and break slowly, reform and repeat the process of bending and breaking.  As $D_\rho/D_E$ is further increased the time scale over which the bands break and reform decreases until bands which span the system no longer form.  Beyond this point structure on a scale which is much smaller than the size of the system can form, and it in some cases these small-scale structures organize into a larger structure as they did in Fig. \ref{StabilityImage}(c).

Identifying the boundary of the stability of the band solution, characterizing the nature of the dynamical states of the system in this regime and comparing to the results obtained for reversing rods \cite{Shi2014} and the nematic Vicsek model \cite{Ngo2013} remain to be done and will be described in future work.

  \section{An Alternative Kinetic Term}\label{iso_limit}

\begin{figure}
  \centering
  \vspace{0.1 in}
  \includegraphics[width=0.8\columnwidth]{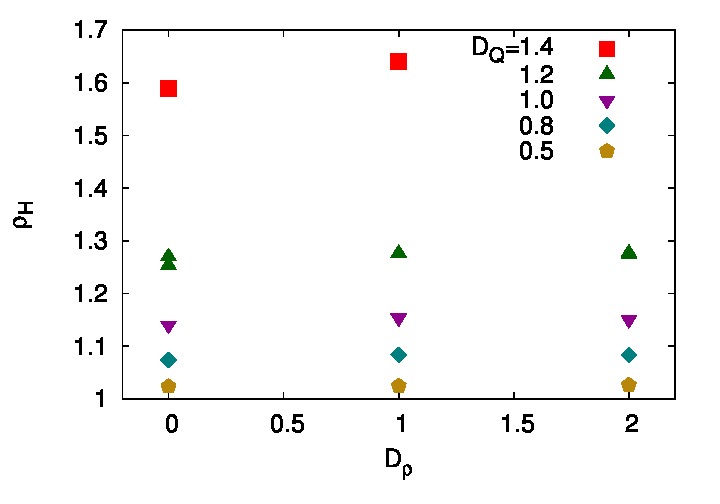}
  \caption{ (color online) The density of the ordered phase ($\rho_H$) is shown for a few different values of the coefficient of the kinetic term ($D_\rho$) in the case where that kinetic term is proportional to $S$. When compared to the plot on the left in Fig. \ref{Insensitive} it can be seen that the change to the kinetic term has not significantly altered the density of the ordered phase. }
  \label{Dr_iso}
\end{figure}

  The theory which has been discussed above includes the kinetic term ($D_\rho$ in Eq. \ref{dtQ1}) which causes any inhomogeneous state to have some local nematic ordering, i.e., the only truly isotropic state of the theory, even when $\rho<\rho_c$ is the homogeneous one.  We can, instead, choose a description in which the theory reduces to that of an isotropic fluid in the limit $S\rightarrow 0$ by replacing $D_\rho$ in Eq. (\ref{dtQ1}) with $S\cdot D_\rho$. This new kinetic term does not change the linear stability of the system, so the effects of this change were investigated numerically with the same method discussed in Section \ref{NumPhen}, and described in terms of the phase separated end state. We found that the final steady states reached by the system are largely the same as those seen in the original theory (also discussed in Section \ref{NumPhen}). In systems with activities and densities ($D_Q$ and $\rho_0$) for which the homogeneous state is unstable, the steady state was a set of bands with the same structure discussed in Section \ref{NumPhen}, and the same phase contrast.  The density of the ordered bands has not changed, which can be seen in Fig. \ref{Dr_iso}, and the density of the isotropic phase is similarly unaltered. Therefore we conclude that the nature of this kinetic term does not affect the phase separation behavior of an active nematic fluid. 

\begin{acknowledgments}
EFP and AB acknowledge support from NSF-DMR-1149266, the Brandeis-MRSEC through NSF DMR-0820492, and the HPC cluster at Brandeis for computing time. EFP also acknowledges support through NIH-5T32EB009419 and IGERT DGE-1068620.
\end{acknowledgments}

% Bibliography
\bibliography{activenematicrefs}

\end{document}